\newcommand{\be}{\begin{equation}}
\newcommand{\ee}{\end{equation}}
\newcommand{\bea}{\begin{eqnarray*}}
\newcommand{\eea}{\end{eqnarray*}}
\newcommand{\ba}{\begin{eqnarray}}
\newcommand{\ea}{\end{eqnarray}}

\documentstyle[12pt]{article}
\title{{\bf Determination of the Higgs Boson Mass by the Cancellation of
Ultraviolet
Divergences in the $SU(2)_L \otimes U(1)$ Theory}}

\author{Gabriel L\'opez Castro \\
Departamento de F\'\i sica,
CINVESTAV del IPN \\
Apartado Postal 14-740, M\'exico  07000, D.F., M\'exico\and
Jean Pestieau \\
Institut de Physique Th\'eorique, Universit\'e catholique de Louvain \\
Chemin du Cyclotron 2, B-1348 Louvain-la-Neuve, Belgium}

\date{}

\begin{document}

\thispagestyle{empty}
\maketitle

\begin{center}
{\large In memory of Roger Decker}
\end{center}

\vspace*{5mm}

\begin{abstract}

We assume the vanishing of the quadratic divergences  in the
$SU(2)_L \otimes U(1)$ electroweak theory. Using $\,$ the $\,$ top $\,$
 mass $\,$ value reported recently
by the CDF Collaboration $m_t = 176 \pm 8 \pm 10 GeV$, we
predict the  mass of the Higgs boson to be
$m_H = 321 \pm 29 GeV$.
If we assume the vanishing of both quadratic and loga-\-rithmic
divergencies of the top self-mass,
we predict $m_t = 170.5 \pm 0.3 GeV $ and $m_H = 308.6 \pm 0.7 GeV.$
\end{abstract}

\newpage
\par\noindent
{\bf 1.} In the $SU(2)_L \otimes U(1)$ theory, quadratic divergences, at
the one loop level, are
universal (i.e. they are the same for all physical quantities). We assume
they vanish \cite{D}-\cite{O}, namely

\vspace*{2mm}
$$
m^2_e + m^2_\mu + m^2_\tau + 3(m^2_u + m^2_d + m^2_c + m^2_s + m^2_t + m^2_b) =
\frac{3}{2} m^2_W + \frac{3}{4} m^2_Z + \frac{3}{4} m^2_H
\eqno{(1)}
$$

\vspace*{5mm}
\par\noindent
In the following we shall neglect the fermion masses other than $m_t$ and
$m_b$.\\
{}From the recent discovery \cite{A} of the top quark, with a mass
\ba
\setcounter{equation}{2}
m_t &=& 176 \pm 8 \pm 10   GeV \nonumber\\
&=& 176 \pm 13 GeV
\ea
obtained by the CDF collaboration at FNAL, we predict from Eq (1), the
Higgs mass
$$m_H = 321 \pm 29 GeV
\eqno{(3)}
$$where we have used \cite{C}
\ba \setcounter{equation}{4}
m_Z &=& 91.1887 \pm 0.0044 GeV \\
m_W &=& 80.23 \pm 0.18 GeV
\ea
\vspace*{3mm}
\par\noindent
{\bf 2.} We assume $m_t$ and $m_H$ to be determined
[1-4,7-9] by requiring the ultraviolet divergencies to vanish for the
(on-mass shell) self-energy of the quark top, namely
$$
\Sigma^{\mbox{div}}_t = - m_t \frac{\alpha}{4 \pi} \; \frac{m^2_Z}{m^2_Z -
m^2_W} F_t (\Lambda) = 0.
\eqno{(6)}
$$
\vspace*{3mm}
\par\noindent
for any $\Lambda \cdot F_t (\Lambda)$, the function containing the cutoff
$\Lambda$ has the general form \cite{H}
\vspace*{3mm}
\ba \setcounter{equation}{7}
&&F_t(\Lambda) \equiv\frac{3}{4} \; \frac{1}{m^2_W m^2_H}\left[ \Lambda^2
\left\{ m^2_H + m^2_Z + 2 m^2_W - 4 m^2_t \right.\right.
\nonumber\\
&&\left.-4 m^2_b \right\}
+ \ln \Lambda^2 \left\{ - \frac{1}{2} m^4_H - m^4_Z - 2 m^4_W
\right.\nonumber\\
&&+ 4 m^4_t + 4 m^4_b + \frac{1}{2} m^2_H (m^2_t-m^2_b) \nonumber\\
&&\left.\left.- \frac{4}{9} m^2_H (m^2_Z - m^2_W)\right\} \right]
\ea

\par\noindent
Eqs(6) and (7) imply the two following relations:
\ba
&&m^2_H + m^2_Z + 2 m^2_W - 4 m^2_t - 4 m^2_b = 0 \\
& & \nonumber \\
&& \frac{1}{2} m^4_H + m^4_Z + 2 m^4_W - 4 m^4_t - 4 m^4_b - \frac{1}{2}
m^2_H (m^2_t - m^2_b)
\nonumber\\
&&+ \frac{4}{9} m^2_H (m^2_Z - m^2_W) = 0
\ea
Of course, Eqs(1) and (8) are identical due to the universality of
quadratic divergences
\cite{D}-\cite{P}. By using Eqs(4), (5), (8) and (9), we predict:
\ba
m_t &=& 170.5 \pm 0.3  GeV \\
m_H &=& 308.6 \pm 0.7 GeV
\ea
The predicted mass for the top quark is in agreement with the central value
reported by the
CDF Collaboration \cite{A}, Eq.(2). We have used $m_b = 5 GeV$.  With $m_b
= 0$, our predictions are
$m_t = 171.1 \pm 0.3 GeV$ and $m_H = 309.7 \pm 0.7 GeV$.\\
Eqs(10) and (11) can be compared with the fitted value of $m_t$  derived
from the
LEP data \cite{C}, $m_t=173 \pm 13 GeV$ if $m_H = 300 GeV$.
Finally, let us mention that the others solutions to Eqs. (8) and (9) given
above namely,  $m_t = 78 GeV $ and $m_H = 58 GeV$,  are excluded by the
present data.
It is interesting to observe that $(2 m_t - m_H) \approx 3(m_Z - m_W)$ and
$m_t \approx m_Z + m_W$.


\vspace*{25mm}


\begin{thebibliography}{90}
\bibitem{D}
R. Decker and J. Pestieau, Preprint UCL-IPT-79-19, Universit\'e de Louvain
(1979); DESY Workshop October 22-24, 1979; Mod. Phys. Lett {\bf A4} (1989)
2733; ibid {\bf A5} (1990) 2579 (E); ibid {\bf A7} (1992) 3773.
\bibitem{V}
M. Veltman Acta Phys. Pol. {\bf B12} (1981) 437.
\bibitem{P}
J. Pestieau, in IIIth Mexican School of Particles and Fields, ed. J.L.
Lucio M. and
A. Zepeda (World Scientific, 1989), p. 281; in The Gardener of Eden, ed. P.
Nicoletopoulos
and J. Orloff, Phys. Mag. {\bf 12} (1990) 193.
\bibitem{O}
P. Osland and T.T. Wu, Z. Phys. {\bf C55} (1992) 569, 585, 593; Phys. Lett.
{\bf B291} (1992) 315; C. Newton, P. Osland and T.T. Wu, Z. Phys. {\bf C61}
(1994) 421, 441.
\bibitem{A}
F. Abe  {\em et al}., CDF Collaboration, {\it Observation of Top Quark
Production in $p\bar p$ Collisions}, FERMILAB-PUB-95/022-E (1995).\\
See also S. Abachi {\em et al}, D$\oslash$ Collaboration, {\it Observation
of the Top Quark},
FERMILAB-PUB 95/028-E (1995), where $m_t$ = 199 $\pm$ 30 GeV is given.
\bibitem{C}
D. Schaile, {\it Precision Tests of the Electroweak Interaction}, in
Proceedings of the XXVII International
Conference on High Energy Physics, 20-27 July 1974, Glasgow, Institute of
Physics Publishing, Bristol and
Philadelphia (1995), p. 27.
\bibitem{Pe}
J. Pestieau and P. Roy, Phys. Rev. Lett. {\bf 23} (1969) 349; R. Decker and
J. Pestieau, Lett. Nuovo Cimento {\bf 29} (1980) 560.
\bibitem{S}
E.C.G. St\"uckelberg, Nature {\bf 144} (1939) 118; A. Pais, Verh. Roy.
Acad. Amsterdam {\bf 19}
(1946) 1; S. Sakata and O. Hara, Progr. Theor. Phys. {\bf 2} (1947) 30;
H. Terazawa, Phys. Rev. Lett. {\bf 22} (1969) 254, 442 (E); Phys. Rev. {\bf
D1} (1970) 2950;
ibid {\bf D4} (1971) 1579.
\bibitem{H}
I-Hsiu Lee and S.D. Drell,  SLAC-PUB-5423, RU 91-5-B (1991); B\'eg Memorial
Volume.
\end{thebibliography}
\end{document}